\documentclass[sigconf, nonacm]{acmart}
\usepackage{csquotes}

\AtBeginDocument{%
  }

\setcopyright{none}

\acmISBN{978-1-4503-XXXX-X/18/06}

\begin{document}

\title{One Does Not Simply ‘Mm-hmm’: Exploring Backchanneling in the AAC Micro-Culture} 

\author{Tobias Weinberg}
\affiliation{
\department{Computer Science}
  \institution{Cornell Tech}
  \city{New York}
  \country{USA}}
\email{tmw88@cornell.edu}

\author{Claire O'Connor}
\affiliation{%
\department{College of Information}
  \institution{University of Maryland}
  \city{College Park, MD}
  \country{USA}}
\email{coconno5@umd.edu}

\author{Ricardo E. Gonzalez Penuela}
\affiliation{%
\department{Information Science}
  \institution{Cornell Tech}
  \city{New York}
  \country{USA}}
\email{reg258@cornell.edu}

\author{Stephanie Valencia}
\affiliation{%
\department{College of Information }
  \institution{University of Maryland}
  \city{College Park, MD}
  \country{USA}}
\email{sval@umd.edu}

\author{Thijs Roumen}
\affiliation{%
\department{Information Science}
  \institution{Cornell Tech}
  \city{New York}
  \country{USA}}
\email{thijs.roumen@cornell.edu}

\renewcommand{\shortauthors}{Weinberg et al.}

\begin{abstract}
  Backchanneling (e.g., "uh-huh", "hmm", a simple nod) encompasses a big part of everyday communication; it is how we negotiate the turn to speak, it signals our engagement, and shapes the flow of our conversations. For people with speech and motor impairments, backchanneling is limited to a reduced set of modalities, and their Augmentative and Alternative Communication (AAC) technology requires visual attention, making it harder to observe non-verbal cues of conversation partners. We explore how users of AAC technology approach backchanneling and create their own unique channels and communication culture. We conducted a workshop with 4 AAC users to understand the unique characteristics of backchanneling in AAC. We explored how backchanneling changes when pairs of AAC users communicate vs when an AAC user communicates with a non-AAC user. We contextualize these findings through four in-depth interviews with speech-language pathologists (SLPs). We conclude with a discussion about backchanneling as a micro‑cultural practice, rethinking embodiment and mediation in AAC technology, and providing design recommendations for timely multi-modal backchanneling while respecting different communication cultures.
\end{abstract}

\begin{teaserfigure}
  \includegraphics[width=\textwidth]{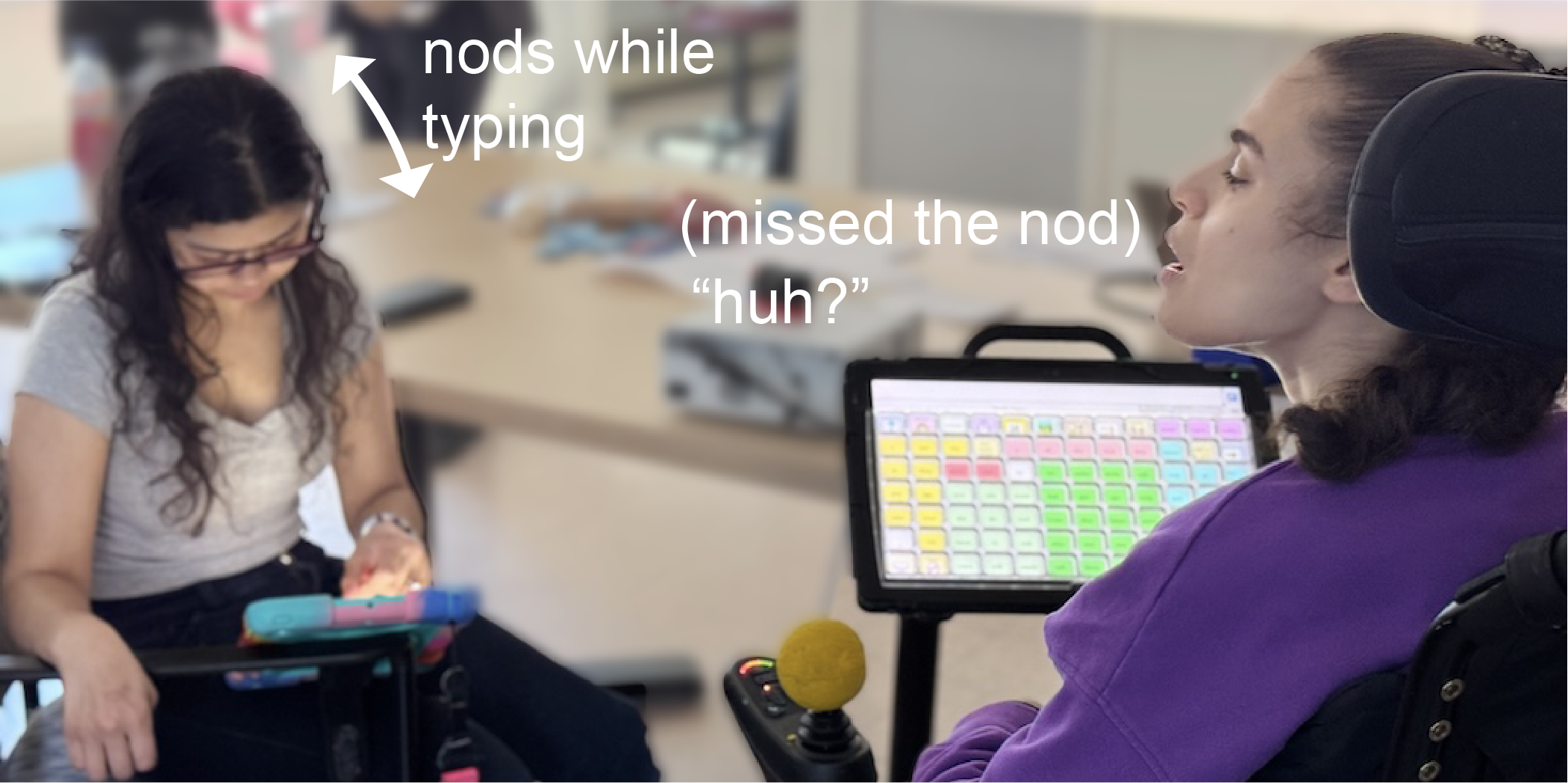}
  \caption{A conversation between two AAC users. We find that backchanneling (active-listening) plays a crucial role in communication. However, here to enter text, both users engage with their devices, missing out on non-verbal cues from their interlocutor. We identify a need for better support of backchanneling for AAC, while respecting their established micro-culture.}
  \Description{Two AAC users are seated across from one another during a workshop activity. Both are looking down at their devices, engaged in composing messages. Despite sitting face to face, their visual attention is directed at their respective screens, not at each other. On the leftmost participant, there is an arrow on top indicating a nod gesture, and on the right participant, a legend that says: "huh?" because she missed the nod.}
  \label{fig:teaser}
\end{teaserfigure}

\maketitle

\section{Introduction}
Everyday conversations are filled with subtle signals that show we care and are listening. A simple nod, a soft “mm-hmm,” or a quick smile can speak volumes without taking the floor~\cite{ward2006non, gardner1997conversation, bavelas2000listeners}. These brief responses—known as backchanneling—are the glue that holds conversations together, reassuring speakers that they are understood and encouraging them to continue. In this paper, in line with Yngve’s definition~\cite{yngve1970getting}, we refer to the term ‘backchanneling’ as the verbal (hmm, yeah, uhu) and non-verbal (nods, hand gestures, body posture) cues that help us negotiate the turn to speak, they serve a meta-conversational purpose, such as signaling listener's attention, understanding, sympathy, or agreement, rather than conveying significant information.  ~\citet{jurafsky1997automatic} showed that in spoken dialogue, backchanneling composes ~19\% of communication.

Users of Augmentative and Alternative Communication (AAC) technology cannot always rely on instant interjections or effortless gestures to engage in backchanneling. These difficulties arise because (1)~there is no clear way to type short backchanneling utterances like “mm-hmm” or express nuance 
 of these using AAC interfaces, (2)~AAC users often focus their visual attention on the device itself, making it hard to observe or respond to their conversation partner’s cues, and (3)~the backchanneling cue may come too late after the topic has already shifted disrupting the flow of conversation. Producing a backchanneling utterance can also interrupt their own message construction, requiring them to pause typing or shift eye focus to tap out a brief response, which can disrupt their communication flow. These constraints limit or render traditional backchanneling modalities limited or out of sync for AAC users, creating barriers to staying actively engaged in conversations.

Despite these challenges, AAC users are far from passive listeners—they have developed creative strategies to stay expressive and involved. In practice, many AAC users become adept multi-modal communicators, combining whatever channels they have available to co-create meaningful messaging~\cite{ibrahim2018design}. We therefore adopt a perspective in which communication is seen as a dynamic, shared endeavor—shaped by common context, embodied signals, and reciprocal engagement—rather than a simple, one-way transfer of unambiguous information when designing AAC technology~\cite{barnlund1970transactional, waller2006communication}. 

We contribute to the broader discourse around expressive and multi-modal communication for AAC technology~\cite{kane_at_2017, valencia_less_2023}. Tools such as \textit{COMPA}~\cite{valencia2024compa} and interfaces developed in \textit{Why So Serious?}~\cite{weinberg2024why} explored ways to address challenges of timing and conversation flow in the context of ongoing conversations involving non-AAC users. Others have explored the role of physical objects to augment non-verbal communication and personal expression for AAC users~\cite{valencia_aided_2021, curtis2024breaking}. In the field of Conversation Analysis, researchers have studied turn-taking in AAC communication; they point out the significance of backchanneling in this exchange~\cite{buzolich1988turn, farrier1985conversational}. However, we still lack a clear understanding of how AAC users navigate backchanneling: how they decide which modality to use, how effective these signals are, and how conversation partners perceive them. Without this understanding, AAC technologies and communication norms risk overlooking a crucial component of expressive communication for people who use AAC. 

In this paper, we explore this gap, asking one overarching question: How can AAC users engage in expressive backchanneling during ongoing conversations using different modalities? We investigate this through a co-design workshop with AAC users~\cite{beneteau2024aligned}, focusing on their lived communication experiences. We contextualize these findings through four in-depth interviews with Speech-Language Pathologists (SLPs). We focus on two research questions:
\begin{enumerate}
    \item What modalities do AAC users currently use for backchanneling, and what are the limitations and opportunities of these methods? 
    \item How do AAC users navigate the dynamics of back-and-forth conversation when backchanneling? In particular, does their approach to backchanneling change based on who they are talking to—e.g., conversing with another AAC user versus with a non-AAC (speaking) partner? 
\end{enumerate}

We found that (1) AAC users develop their own specific micro-culture of communication when it comes to backchanneling, using a wealth of multi-modal cues. When augmenting such communication, it is important to respect that existing micro-culture instead of imposing generic cultural norms. (2) The multi-modal nature of backchanneling reveals a need for embodied interaction with AAC technology. Rather than solely relying on users typing or tapping buttons, they communicate using other modalities, which could empower backchanneling. (3) Our SLPs all indicate the importance of backchanneling, yet realize they are not educated to teach backchanneling, and thus their clients miss out on this important form of communication. There is a need for a more holistic and expressive curriculum. And finally, (4) we derive recommendations for the design of future AAC technology. 

By understanding how AAC users express attentive listening, we can design inclusive communication interfaces to support natural conversation flow for people of all abilities. Treating AAC users as multi-modal communicators helps “level the playing field” in conversation design, ensuring that features for feedback and turn-taking accommodate a slower pace or alternative signals. Different cultures have unique backchanneling norms~\cite{li2006backchannel, white1989backchannels, cutrone2005case} and listeners must learn to bridge those differences, so conversations between AAC and non-AAC users require mutual adaptation and understanding, instead of naively forcing non-AAC norms onto AAC users through the design of technology.
\section{Positionality Statement}
The lead author of this paper has a speech disability and relies on AAC technology. To highlight how this perspective influenced the study, we disclose his experiences, as they played a significant role in shaping the research process.

The lead author has been an AAC user for 13 years since a neuro-motor disease manifested when he was a teenager, limiting his speech and movement. "I had to learn to adapt my communication style to this new form of communication." This experience inspired his research on how to increase expressivity for AAC users.

Backchanneling is essential for any communication, but particularly hard for AAC users. "When I talk with someone, I like to keep eye contact to remain engaged in the conversation, but I can't type my message without looking at my screen". In society, when one is looking at their phone and typing while another person is speaking, it's a clear sign of disengagement in the conversation: "Since I use my phone as AAC, people who don't know me might think that I'm not listening while I'm typing". To compensate for this he would use his eyebrows and facial expression while typing to show that he is actively listening "I think you can get a lot of what I'm thinking just from facial expression, I was always pretty expressive with my face almost clownesque but since I got sick I rely much more on that".

Understanding the different ways that AAC users use backchanneling is crucial to designing tools that can aid in this task: "Backchanneling is a cultural agreement on communication. If you move to another country, that prior 'cultural agreement' you were used to might change. AAC users form their own micro-cultural agreement on communication, and we need to learn how that micro-culture works." We hope this leads to the design of tools that support them without overriding them with the non-AAC cultural agreements.

\section{Contribution, Benefits, and Limitations}

We explore how AAC users employ \emph{backchanneling}—subtle cues such as eyebrow raises, chair taps, and \textit{QuickFire} phrases—to negotiate the perennial trade‑offs of \emph{timing, effort, and modality} to sustain conversational flow. Our study combines a co‑design workshop with AAC users and follow‑up interviews with speech‑language pathologists. We make the following contributions:

\begin{enumerate}
  \item We provide a multi‑modal account of how AAC users in our workshop engage in backchanneling, documenting their listener‑feedback strategies.
  \item We contextualize these observations with in-depth interviews with Speech-Language Pathologists, further highlighting the role of backchanneling in SLP practice. 
  \item  We distill our findings into recommendations to include backchanneling into the design of future AAC technology.
\end{enumerate}

The study (1)~foregrounds AAC users’ lived expertise, offering designers of technology useful insights to support backchanneling; and (2)~equips speech‑language pathologists with empirically grounded talking points and training examples to extend pragmatic‑language curricula.
 
Limitations include that our participant pool was small (four AAC users, four SLPs), English‑speaking, and recruited from a single metropolitan region; as such, our findings should be interpreted as exploratory rather than generalizable. The first author’s dual role as researcher and AAC user enriches but also colors the analysis; we mitigated this through multi‑researcher coding and having multiple observers during the workshop.

\section{Related Work}

We build upon work on challenges in AAC and expressiveness, coordinating feedback in AAC conversations, and cross‑cultural Perspectives on Backchanneling.

\subsection{Challenges in AAC and Expressiveness}
Most AAC users communicate at a rate of 12–18 words per minute (wpm), compared to 125–185 wpm for typical speakers~\cite{waller_telling_2019}. This asymmetry in communication disrupts the flow and creates barriers to effective communication~\cite{higginbotham2013slipping}. \citet{higginbotham2016time} studied how mismatched timing influences the perceived competence of AAC users: when an AAC device output lags behind the conversational flow, partners view AAC users as less capable or disengaged. \citet{clark_grounding_1991} found that delayed response time to form messages takes attention away from the speaker's utterances, making it difficult for AAC users to maintain the conversation.

However, timing is not the only factor shaping communicative success. Communication is not merely the transfer of information from sender to receiver, but a dynamic process in which meaning is co-constructed through shared context, embodied cues, and reciprocal understanding~\cite{barnlund1970transactional, waller2006communication}. Within this model, \textit{conversational agency} (an individual's capacity to express and achieve their goals in conversation) is not solely a function of speed or clarity, but emerges through interaction with social, material, and temporal structures~\cite{valencia_conversational_2020}. Agency is shaped by social constraints, the pace of conversation~\cite{kane_at_2017}, and the availability and responsiveness of communication resources, including vocabulary~\cite{demmans2012towards}, AAC technology, and the context of interaction~\cite{curtis2022state}. As such, supporting expressive backchanneling in AAC is not simply about reducing delay, but about enabling users to participate meaningfully in the construction of mutual understanding.

Expressivity in AAC is governed by these competing goals: agency vs timing, although in most cases AAC users will prioritize agency~\cite{valencia_less_2023}, \citet{weinberg2024why} found that in time pressured scenarios, like making timely humorous comments in an ongoing conversation, this priority shift towards timing and AAC users are willing to trade in agency to deliver the comment faster. These findings highlight the complex dynamics that AAC users navigate during a conversation. 

To address existing asymmetries in timing and agency across AAC users and their interlocutors; researchers have increased AAC users' communication rates by incorporating LLM-based suggestions~ \cite{shen_kwickchat_2022, valencia_less_2023, yusufali2023bridging, cai_using_2023, fang_socializechat_2023, hackbarth2024revolutionizing}, using visual output~\cite{fontana_de_vargas_aac_2022, fontana_de_vargas_automated_2021}, wearable devices~\cite{curtis2023watch},  or even physical attributes ~\cite{bircanin2019challenges, curtis2024breaking} to support the conversational flow~\cite{valencia2024compa,sobel2017exploring,fiannaca2017aacrobat}. However, these discussions often treat backchanneling as a secondary or implicit issue. In contrast, our work centers backchanneling as the primary object of study, offering a detailed, interactional account of how AAC and non-AAC partners coordinate presence, responsiveness, and conversational rhythm. By focusing specifically on the role of backchanneling in mediating timing and agency, we contribute a deeper understanding of the expressive practices AAC users engage in and how breakdowns in these practices shape conversational dynamics. By understanding how AAC users approach backchanneling, we can design inclusive communication interfaces, as well as contribute to existing guidelines for developers~\cite{martin2024bridging} to support natural conversation flow for people of all abilities.

\subsection{Coordinating Feedback in AAC Conversations}
In face-to-face communication, listeners routinely use gestures, facial expressions, and eye contact to signal engagement and understanding~\cite{clark_grounding_1991, fulcher2019interacting, higginbotham2002aac}. These cues serve as immediate and often subtle forms of feedback, allowing for smooth turn-taking without interrupting the speaker~\cite{binger_effects_2008}. Conversation Analysts have examined how turn-taking unfolds in AAC-based interactions, highlighting the essential role that backchanneling plays in these exchanges~\cite{buzolich1988turn, farrier1985conversational}. However, AAC conversations introduce distinct coordination challenges: users must visually attend to their devices to compose messages, and many experience motor constraints that limit their ability to deliver quick listener feedback~\cite{ beukelman_augmentative_2020, valencia_aided_2021}.

Talk-in-interaction is a topic in conversation analysis focused on seamless turn-taking in conversational flow. Researchers in this domain have shown that communicators dynamically co-construct feedback strategies depending on their role, the medium, and the context of interaction~\cite{clark1996using, clarke2013aac, higginbotham2002aac, higginbotham2013primer}. \citet{clark_grounding_1991} had established this notion as achieving "common ground" in which both communication partners come to a mutual agreement in the conversation. In AAC communication, \citet{seale2020interaction} found that AAC users often find it difficult to achieve "interaction symmetry" as AAC users don't have the same means of communicating as quickly as oral speakers. This creates a gap in opportunities for co-constructing feedback responses between the AAC user and their conversational partner. 

\citet{ibrahim2023common} analyzed how children who use AAC coordinate communication with teachers in multi-modal classroom interactions. Their work highlights how users blend gestures, gaze, vocalizations, and devices to collaboratively establish common ground. 

Prior work has explored how AAC users manage turn-taking asymmetries and timing mismatches through strategies like rephrasing, pre-playing utterances, or relying on partners to scaffold the interaction~\cite{bloch_understandability_2004, tegler_aided-speaking_2024}. \citet{seale2020interaction} found that AAC devices might not effectively support individuals in situations requiring active direction giving and information sharing. These findings reveal a persistent tradeoff: AAC users must often choose between showing that they are listening or composing their next contribution.

This tradeoff is especially pronounced in backchanneling, where listener cues must be both timely and subtle. Non-AAC partners may miss these cues or prematurely interrupt, assuming the user has nothing more to add~\cite{legel2025self, higginbotham2016time}. 
\citet{valencia_conversational_2020} study on user agency quantified and categorized non-verbal gestures as meaningful contributions to conversation, highlighting how the use of these gestures supported the agency of AAC users and how they are sometimes overlooked.
While experienced communication partners can interpret signals such as eyebrow raises or body posture~\cite{kraat1987communication}, this skill is not universal. As \citet{doak_rethinking_2021} notes, building familiarity with an AAC user's expressive repertoire is essential to achieving balanced interaction.
To mitigate these gaps, \citet{valencia_co-designing_2021} proposed co-designed “Sidekicks”— companions to amplify an AAC user's nonverbal signals to help them re-enter the flow of conversation. Other work has emphasized the power of multi-modal feedback—combining gestures, sounds, and physical objects—to help AAC users maintain communicative intent while composing speech~\cite{valencia_aided_2021}.
Previous work has demonstrated that by programming insults into a switch, teenage AAC users were able to manifest the voice of a teenager with an attitude~\cite{hornof2009designing}, which showcases the opportunity to use physical switches for more expressive interactions.
The above-mentioned works make impactful contributions to AAC communication, which naturally has implications for backchanneling (or the lack thereof). In this work, we take a specific lens to zone in on the intricacies of backchanneling itself. Our study builds on this by examining how AAC users coordinate backchanneling specifically: how they signal active-listening, manage timing tradeoffs, and construct feedback strategies across modalities.

\subsection{Cross‑Cultural Perspectives on Backchanneling}
Backchanneling behaviors are particular to language and culture. This has been studied for many decades~\cite{yngve1970getting, kjellmer2009we, white1997back, orestrom1983turn}. Previous research has studied backchanneling in multicultural settings~\cite{li2006backchannel, white1989backchannels, cutrone2005case}, for bilingual speakers~\cite{heinz2003backchannel, tao1991english, kubota1991use}, and also the impact in human-robot interactions~\cite{park2017telling, fujie2004conversation, murray2022learning, park2017backchannel}.  
In American English, verbal backchanneling usually takes the form of non-lexical sounds (e.g., hmm, uh-huh, aham)~\cite{ward2006non}. \citet{senk1997analyzing} defined a list of six functions of backchanneling: (1)~continuer, (2)~understanding, (3)~support~and~empathy, (4)~agreement, (5)~emotive, and (6)~minor~additions. In line with this work, we analyze how AAC users engage in backchanneling within face-to-face conversations~\cite{bertrand2007backchannels, truong2011multimodal}. Previous researchers have highlighted the importance of backchanneling for AAC users~\cite{todman2008whole}; little is still known about the role of backchanneling across modalities~\cite{friginal2013linguistic}. Building on \citet{hubscher2023role}, who conducted a multi-modal analysis of discourse markers in AAC interviews, we treat AAC communication as a micro‑culture and compare backchanneling in AAC–AAC versus AAC–non‑AAC dyads to reveal where AAC backchanneling practices diverge from or align with mainstream conversational norms. These insights can guide the design of AAC systems that are sensitive to cultural differences in how backchanneling is expressed and interpreted.
\section{Co-Design Workshop: The AAC Backchanneling Micro-Culture}

To better understand how AAC users engage in backchanneling, we ran a co-design workshop at a local day-care facility. We recruited AAC users and person-centered technology specialists from YAI (partner organization) to explore the role of backchanneling as an AAC user and the potential for technology to support this.
Our institutional review board deemed the study exempt from IRB review, protocol number IRB0148755.

\subsection{Participants}

We recruited four AAC users from our network and two person-centered technology specialists (S1, S2) from our partner organization. Two of them are part-time AAC users (P3, P4) (they can rely on natural speech), while the other two are full-time AAC users (P1, P2). Table~\ref{tab:participants} shows the demographics of the participants. All participants also had motor disabilities ranging from mild to profound disabilities. We categorized them:  Mild (minor limitations in movement), Moderate (noticeable limitations, performs daily activities with some assistance), Severe (requires assistance for most daily activities), and Profound (unable to perform daily activities without assistance).

\begin{table*}[h]
    \centering
    \resizebox{\textwidth}{!}{
        \begin{tabular}{lllllll}
             \textbf{ID} & \textbf{Age} & \textbf{Gender} & \textbf{Motor Disability} &\textbf{AAC Use}& \textbf{AAC Device/s} & \textbf{Input Modality} \\
             P1 & 25-34 & Female & Profound & Full-time & Accent1400, Nuvoice & Direct touch with guard \\
             P2 & 25-34 & Female & Mild & Full-time & Phone/iPad  & Direct touch \\
             P3 & 25-34 & Male & Severe & Part-time & Phone & Direct touch \\
             P4 & 35-44 & Female & Profound & Part-time  & Phone/iPad & Direct touch \\         
        \end{tabular}
    }
    \caption{Participant details}
    \Description{This table presents details of seven participants, including their age, gender, severity of motor disability, AAC devices used, and input modalities.
P1: A 25-34 year-old female with profound motor disability (requiring significant assistance for daily activities), using an Accent1400 with Nuvoice and direct touch input with a guard.
P2: A 25-34 year-old female with mild motor disability (minor movement limitations), using a phone/iPad with NovaChat/TouchChat and direct touch input.
P3: A 25-34 year-old male with Severe motor disability (requires assistance for most daily activities), using a Phone with direct touch input.
P4: A 35-44 year-old female with profound motor disability, using a phone/iPad with direct touch input.
}
    \label{tab:participants}
\end{table*}

\subsubsection{Data Collection and Analysis}
Data from the workshop was collected via video and audio recordings, and three researchers took observational notes during the workshop. They paid attention to the differences between AAC and non-AAC users in producing backchanneling utterances, and they time-stamped key moments of interactions. Researchers debriefed immediately after the workshop to discuss and address any disagreements in observed behavior. Two researchers, who were in attendance at the workshop, reviewed the recordings and took notes independently, labeling interactions and identifying recurring patterns related to the production, reception, and timing of backchanneling. Through iterative discussions, these observations were clustered into broader themes that captured key insights across participants. This collaborative process ensured that the final themes were grounded in the data and reflective of diverse perspectives within the team.

\subsection{Procedure}
The workshop took 3 hours, and it was divided into two activities, each spanning around 1-1.5 hours with breaks in between. One facilitator presented the concept of backchanneling with video examples, outlined the goal of the workshop, and introduced two activities: an exploratory activity and a performance-based activity. In the second activity, AAC users were broken up into two pairs of AAC and non-AAC users and one pair of two AAC users.

\textbf{Activity 1: modalities used in backchanneling. }  
Participants explored the different modalities they currently use to express backchanneling cues. The facilitator prompted reflection by asking questions that addressed facial expressions and eye gaze, gestures and body movements, vocal and sound-based cues, haptic and tactile feedback, and AAC-specific features. The facilitator made sure to give each participant a turn to answer at least one question and extra time to allow them to respond.

\textbf{Activity 2: informal conversation with different speaking partners. }  
Participants engaged in live conversation. The activity began with a brief introduction to types of backchanneling: (1)~continuer, (2)~understanding, (3)~support~and~empathy, (4)~agreement, (5)~emotive, and (6)~minor~additions~\cite{senk1997analyzing}.  We used the examples from~\citet{cutrone2010backchannel} in short videos recorded by one of the researchers. Participants engaged in dialogues in pairs, alternating between AAC-to-AAC and AAC-to-non-AAC contexts. Each pair was assigned one observer-researcher. Participants were encouraged to experiment with timing, modality, and expressivity by discussing daily activities, hobbies, and their background. Each conversation took 20 minutes. Participants then engaged in group discussion on how backchanneling flowed, where breakdowns occurred, and what strategies supported effective communication.

\subsection{Results}

We break the results down into the two activities.
\\\\
\noindent\textbf{Activity 1: Exploring Modalities used in Backchanneling.}

The main themes identified in this activity were: novel opportunities for backchanneling, the constraints of timing and mobility on backchanneling, and collaborative backchanneling. Here we discuss observations that lead to each of these themes.

\subsubsection{Opportunities for Backchanneling}

All participants shared examples of backchanneling they use during everyday conversation. Common examples included facial expressions, nodding their head~(P1, P2, P3), and using certain gestures such as sign language~(P1, P2) to convey messages: “I use my hands [to sign] for help” (P2). Other unique modalities discussed included P1 saying, "I think I sigh a lot, when I'm tired or frustrated," and P4 saying, "I use my chair, I stand up [raise the chair]" to express interest. 

\begin{figure}
    \centering
    \includegraphics[width=1\linewidth]{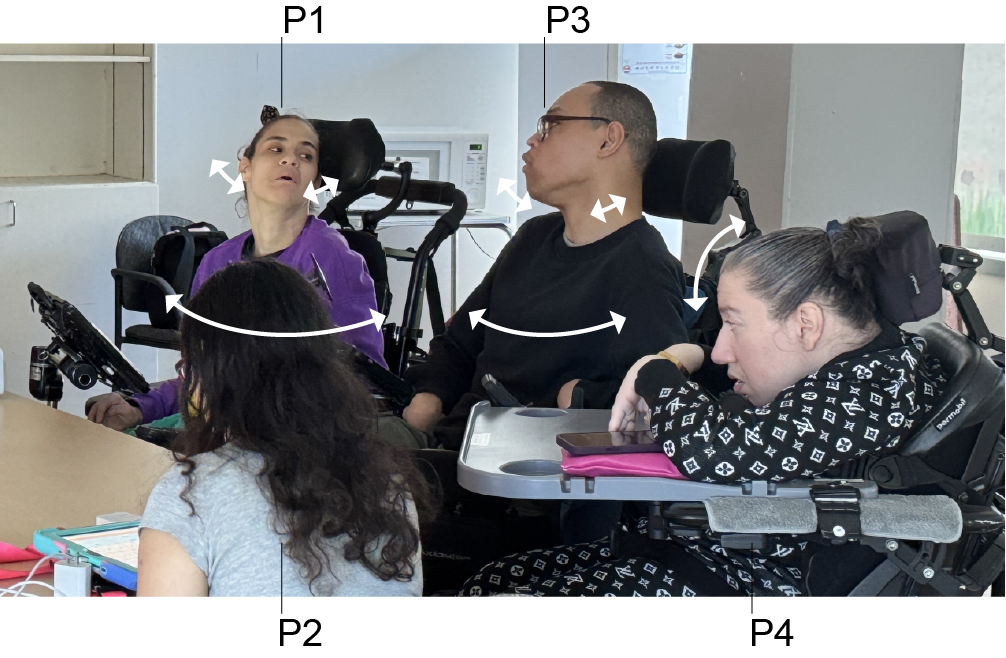}
    \caption{P3 and P1 always shifted their bodies while P4 only moved her head in order to face their communication partners.}
    \Description{"Four AAC users sitting in a circle during a workshop activity. Each participant is engaged in conversation and displaying different physical cues. P3 and P1 on their powered wheelchairs visibly shift their upper bodies toward their partners, while P4, seated in a powered wheelchair, tilts her head slightly. P2 sits upright with attentive posture. The image captures the diversity of nonverbal backchanneling behaviors, such as body orientation and head movement, used to signal engagement across different mobility levels."}
    \label{fig:shift-body}
\end{figure}

When observing other backchanneling among participants, head and body movements varied based on the individual's abilities. For instance, as we can see in Fig.~\ref{fig:shift-body}, P3 and P1 always shifted their bodies while P4 only moved her head in order to face their communication partners. Eye gaze also varied as some of the participants were either focused on preparing their message (P1 and P2) or adjusting their chair (P3). Out of all participants, P2 maintained eye contact for most interactions. 
In addition to nonverbal backchanneling, all participants also exhibited verbal backchanneling, such as P1 saying "yeah" and P4 saying "yes" when confirming their understanding of the message. P3 exhibited unique verbal backchanneling behaviors when saying "one sec" to the non-AAC user when indicating that they needed to adjust their chair. 

\subsubsection{Constraints in Timing and Mobility Shape Backchanneling Opportunities}

A challenge for participants was aligning backchanneling with the fast pace of spoken conversation. AAC typing requires visual and motor focus, leading to delays or missed opportunities to use backchanneling. For example, P2 asked P1 "what do you want to do besides painting?" to which P1 played a pre-programmed message on her AAC device, listing hobbies "... and I love to go wheelchair hiking", P2 reacted by doing a thumbs up in support, adding "I went wheelchair hiking last year". Since P1 played a pre-programmed message, her focal attention was on P2 for the backchanneling reaction, but after P2 answered, she engaged with her device, missing P1 saying "me too, where?"

Physical access to devices complicated real-time signaling. P4, who needs assistance to set up her iPad, said: “I can’t reach it,” prompting P2 and the staff to adjust her device. Similarly, P1 had trouble indicating confusion during an explanation: she thought she was backchanneling that she was thinkingthe facilitator missed that, because the pace of conversation prevented her from composing an elaborate message like "give me a second". Which in turn forced her to type it out, falling back to verbal language, “I’m thinking, I don’t understand the question,” instead of nonverbal signaling later attempted to comment on a joke by P3; however, before she had produced her response, the facilitator moved on, which led her to stop typing. 

These experiences underscore how the duration of composition and the motor effort required to access the device limit timely backchanneling. As a result, participants occasionally found themselves locked out of a conversation’s rhythm, breaking flow. P1 explicitly noted, “I feel like AAC doesn’t help me show that I’m listening. I’m more physical, I feel like,” suggesting that she relies on facial expression to avoid these device-related timing pitfalls.

\subsubsection{Nonverbal Signals Offer Efficient Backchanneling}

Participants leaned on nonverbal cues, as quick alternatives that can operate “in parallel” to device use. These cues included head nods, facial expressions, eyebrow movements, and even small chair sounds, all of which allowed users to remain active listeners without interrupting their typing.

P1 offered a rationale: “I feel like my face shows how I’m feeling.” By raising her eyebrows, lowering her head, or widening her eyes, she could immediately reflect her reactions. P3 echoed this, remarking, “When I’m mad/frustrated, I have a tendency to bite my lower lipnd, P2 described how she uses eyebrows for acknowledgment, then smiled and shrugged: “I just like moving my eyebrows.”

Participants also used body-based signals that filled the “listening gap” while focusing on the device, for example, leaning forward or moving their head. For instance, P1 waved her arm back and forth while saying "wait" that she was typing.

Another use of nonverbal signals was P1 making a "beep" noise from her chair to signal she was stepping away to get water, drawing attention without forcing ae. P3 expressed: "When I'm frustrated or pissed of I would bang my arm on the armrest" showing ways they use their chairs as a channel for backchanneling.

All of these illustrate how nonverbal channels AAC users to show “I’m here, I’m following” without interrupting their composition flow.

\subsubsection{Collaborative Communication and Shared Backchanneling}

Participants co-constructed meaning by clarifying utterances from others; backchanneling can be collective rather than strictly a one-to-one exchange. This was prevalent when P4 used her dysarthric voice: P2 and P3 often acted as interpreters, when P3 explained, “She uses her body, and she calls me on FaceTime, or text,” to elaborate on P4’s partially audible statement. P2, P3, and P4 already knew each other from attending the same day hab program at the facility. P2 and P1 knew each other from a social event that took place at the facility prior to the workshop. 

Group humor fueled collaborative backchanneling. When P3 crashed his chair into a table, P2 provided a playful real-time commentary—“car crash… take his license away”—eliciting laughter from everyone. P4 later verbally responded to a question, but her words were unclear; P2 jumped in to clarify “emojis,” referring to a facial reaction, and the rest of the group agreed they use emojis as reactions in text.

These interactions reveal a “community-based” approach to maintaining conversational flow, where the communication task is distributed among participants who "share the load". If one AAC user encounters a breakdown—due to timing, clarity, or device access—others fill the gap.
\\\\
\noindent\textbf{Activity 2: Informal conversation with different speaking partners.}

We found variations in strategies for backchanneling between AAC users or when there was an AAC user and a non-AAC user. Here, we break down these observations.

\subsubsection{AAC–AAC Interaction and Backchanneling}
Our observations underscore the distinctive ways in which two AAC users (P1 and P2) navigated conversational turns, demonstrated listening, and offered backchanneling cues. Although both participants were familiar with AAC devices, their strategies for signaling engagement varied widely and often involved subtle, multi-modal behavior.

\begin{figure} [h]
    \centering
    \includegraphics[width=\linewidth]{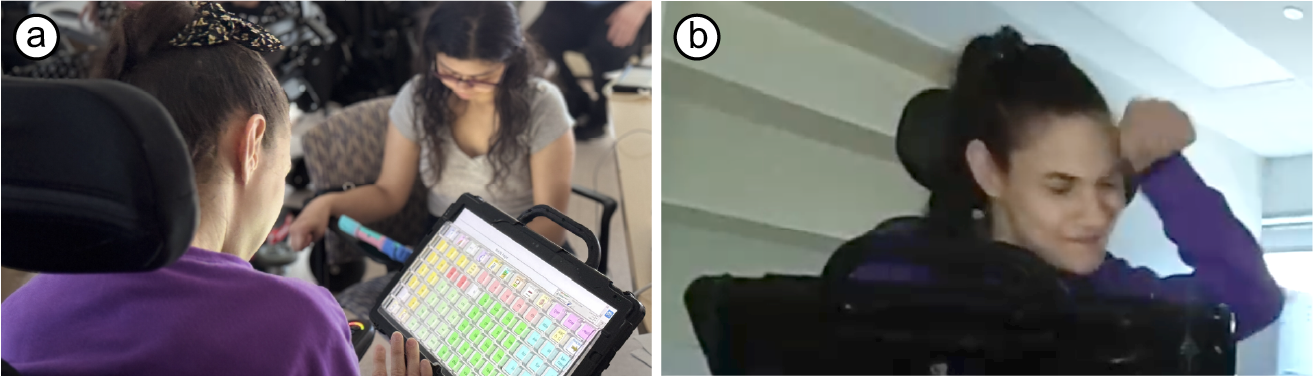}
    \caption{(a)~both P1 and P2 typically looked downward while entering text. (b)~P1 playfully tapped her forehead (“duh!”), then typed, “I knew that.”}
    \Description{This is a 2-panel figure in (a) A person using an AAC device with a color-coded grid layout is engaged in conversation but looking at her device. She is seated in a wheelchair, and her conversation partner, also seated, looks down attentively at her own AAC device.
    In panel (b), the same AAC user from panel (a) is smiling and making an expressive forehead tap gesture, commonly used to signal “I knew that!” The gesture is captured mid-motion, reflecting a playful and emphatic listener response.}
    \label{fig:eye-contact}
\end{figure}

\textbf{Reduced Eye Contact During Composition:}
One characteristic of AAC–AAC communication was the near-complete loss of eye contact when either participant was typing. Because composing a messagends visual attention to the screen, as shown in Fig.~\ref{fig:eye-contact}a, both P1 and P2 looked downward while entering text. When P1 asked P2 follow-up questions (e.g., “What kind of jewelry do you do?”), it went initially unnoticed by P2 because she was composing a message replayed messages, and only then did P2 acknowledge by typing her response. These repeated prompts highlight the difficulty inherent in AAC-based backchanneling: the listening partner may not notice if they are focused on their device. This breakdown occurs mostly for visual backchanneling. To illustrate this, there was an instance where P1 was typing and P2 made a thinking gesture [like thinking what to ask next], but P1 missed the gesture. This example illustrates a fundamental breakdown in communication when backchanneling is missed, since AAC users focus their visual attention on the device to compose messages.

\textbf{Gestural and Nonverbal Cues:} Despite the reduced eye contact, both participants employed nonverbal signals—some unintentional and others very deliberate—to convey engagement. For instance, immediately after P2 said “I like to do jewelry,” P1 playfully tapped her forehead (“duh!”), shown in Fig.~\ref{fig:eye-contact}b, then typed, “I knew that.” Although the verbal content arrived with a delay, the forehead tap itself acted as an immediate acknowledgment. P2 also raised her head to look for confirmation from P1 once she finished playing a message, effectively turning a simple head-lift into a question: “Did you hear me?” or “What do you think?” Occasionally, they punctuated typed responses with a quick nod or thumbs-up—a rapid indicator that does not require prolonged eye contact.

\textbf{Timing Mismatches and Overlapping Composition:}
A recurring pattern was simultaneous composition. On several occasions, P1 and P2 started typing simultaneously, unaware that the other was typing too. This dynamic mirrors two spoken conversationalists talking over each other, but in the AAC context, it produces extended silences. In one example, P1 finished typing—“How long have you been doing it?”—just as P2 played a different message: “I went horseback riding yesterday.” Both talk turns emerged back-to-back, with neither referencing the other’s topic. Although each participant offered verbal or gestural backchanneling (e.g., “Ohh,” from P1 in response to the horseback-riding), their prior question about jewelry went unanswered, and the conversation moved on.

\textbf{Leveraging Auditory Backchanneling:}
Although typing was the primary mode of communication for P1, she frequently complemented quick verbal reactions as backchanneling cues, particularly when she wanted to respond swiftly. For instance, upon learning that P2 had gone horseback riding, P1 responded with an enthusiastic “Ohh!” followed by typing  “where?” rather than pausing to compose those sentiments on her device. Similarly, when P2 shared one of her poems, P1 punctuated her appreciation with a bright “wooow,” conveying admiration without losing the conversational thread. These quick verbal exclamations helped P1 stay engaged in real time, showing that she was actively following along even when longer typed responses might have lagged behind the ongoing conversation.

\subsubsection{AAC to Non-AAC Interaction and Backchanneling}
Our observations of two AAC–to–non‑AAC pairs (P3–S1 and P4–S2) reveal a characteristic backchanneling pattern that differs from AAC‑to‑AAC exchanges. 
Three analytic threads stand out: confirmation by repetition, temporal adaptation, and multi-modal supplementation.

\textbf{Confirmation by Repetition the Default Backchanneling:}
Both non‑AAC partners relied on echoic backchanneling cues, repeating the AAC user’s last few words with rising intonation to solicit confirmation. When P3 explained his plan to return to college, S1 continually mirrored fragments of P3’s dysarthric speech:
\begin{displayquote}
S1: “…go back to college?”

P3: “Yeah, yeah.”
\end{displayquote}

A similar pattern emerged when P3 listed video‑game titles “…Need for Speed”—prompting S1’s clarification “the racing game?” and P3’s rapid “yeah, yeah.” This backchanneling (repetition as question) served two functions: it verified intelligibility and explicitly handed P3 the conversational floor to continue.

A similar pattern emerged in the P4–S2 pair, where P4’s dysarthric speech was consistently echoed by S2 for confirmation. For instance, P4 remarked, “I would like to go to college for a poetry class.” S2 repeated, “Oh, college poetry nice,” confirming she had partially understood. Each time S2 restated P4’s words, P4 answered with a brief “yes” or a nod, thereby validating S2’s comprehension and signaling readiness to proceed.

\begin{figure}[h]
    \centering
    \includegraphics[width=1\linewidth]{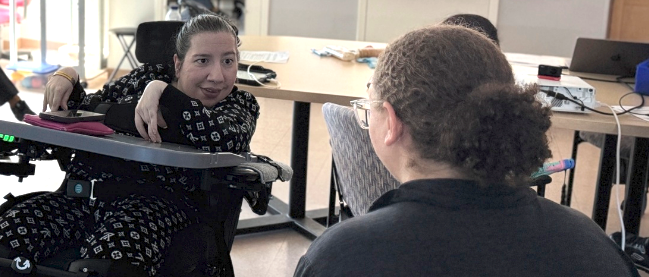}
    \caption{P4 maintained continuous eye contact and responded with tiny nods.}
    \Description{"P4 seated in a wheelchair is leaning slightly forward with her arm resting on her tray, making direct eye contact with her conversation partner. She appears engaged and expressive, smiling gently and responding with tiny nods as she listens. Her partner is seated across from her."}
    \label{fig:P4-eye-contact}
\end{figure}

S2 also used frequent short backchanneling (“aha,” “yeah,” “okay”) much like S1 did, but often inserted them after almost every phrase, reflecting how P4’s speech required incremental checks. Despite this frequency of backchanneling, P4 maintained a continuous eye contact and responded with “very tiny nods” each time S2 mirrored her words as shown in Fig.~\ref{fig:P4-eye-contact}. The sequence of speak–repeat–confirm created a rhythmic back-and-forth, minimizing misunderstandings.

\textbf{Evolving Backchanneling Frequency:} S1 peppered short acknowledgments—“yeah,” “gotcha,” “okay”—throughout the dialogue, especially early on, when he was still learning P3’s speech patterns. As the conversation progressed and S1 became more comfortable, the frequency of these utterances decreased. This change suggests that frequent backchanneling can be critical for establishing a “listening baseline” and then naturally recedes once mutual understanding is achieved. When S1 asked if P3 had used noise-canceling headphones, he used minimal backchanneling like “hmhm” encouraging P3 to continue without interrupting the flow of P3’s story.

Similarly, S2 adapted her listening approach over the course of the exchange. Early on, she repeated longer segments to verify correctness, but later she used fewer, more targeted prompts. For example, when P4 mentioned she liked Broadway shows, S2 asked a simpler question—“Have you watched [Wicked]?”—to which P4 affirmed with her head. Meanwhile, this topic also invited an interjection from P2 (across the room) who overheard “Wicked,” chimed in with a head nod and added via AAC, “it’s on Peacock,” then seamlessly returned to her own conversation.

Over time, the pairs co-constructed a smoother, less backchannel-heavy style of interaction, relying more on natural give-and-take and less on constant repetition.

\begin{figure}[h]
    \centering
    \includegraphics[width=1\linewidth]{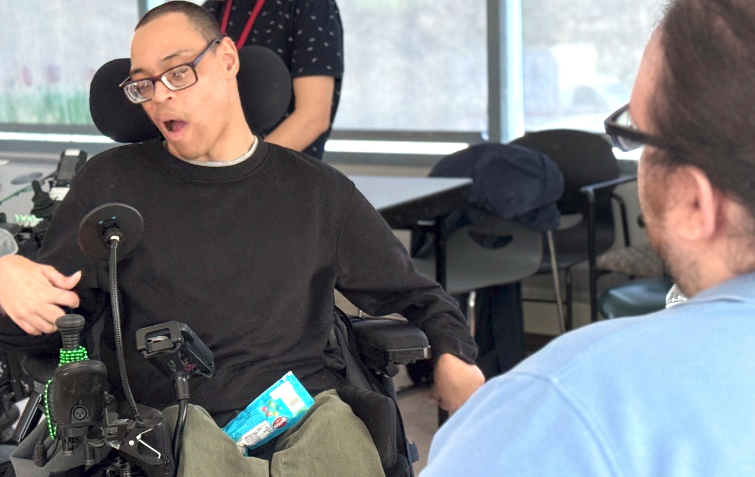}
    \caption{P3 would look to the side each time he is thinking.}
    \Description{An AAC user in a motorized wheelchair is mid-conversation with a speaking partner seated across from him. He is looking to the side, with his mouth slightly open in speech. His body is angled slightly to the side,  indicating emphasis or thought.}
    \label{fig:p3-look-side}
\end{figure}

\textbf{Multi-modal Supplementation and Affective Nuance:}
Although speech carried the bulk of meaning, both AAC users layered nonverbal cues onto their verbal acknowledgments. Fig.~\ref{fig:p3-look-side} shows: P3 “would look to the side each time he is thinking,” signaling floor‑holding while composing his next utterance, and P4 maintained “almost constant eye gaze… with very tiny nods” to ratify S2’s echoes. Emotional subtleties were also conveyed multi-modally: when S2 offered help finding an online poetry class, P4 “raised her eyebrows like ‘really?’ with a big smile,” then voiced “yes.” Later, S2’s travel anecdote elicited P4’s elongated “yeeeh” accompanied by a broad grin, demonstrating that emotional feedback is supported by facial expressions.

\subsubsection{Concluding Group Discussion}
At the end of the session, we concluded with an all-hands group discussion. Participants described supportive collaboration as key to sustaining effective AAC–to–Non-AAC exchanges. P3 explained, “Sometimes when I have a hard time, [P4] would help me, or the other way around,” illustrating how they jointly navigated communication breakdowns. Even near the end, P4 asked S2 to hand out business cards mid-conversation, prompting S2 to promise assistance and P4 to respond, “You are the best.” Such moments highlight how cooperative problem-solving (e.g., clarifying phrases, physically assisting one another) remained central to their communication success.

Another theme was varying backchanneling styles based on familiarity. When asked if backchanneling changes with close family or romantic partners, everyone confirmed “yes” (by voice or head nod), suggesting they adapt their acknowledgment cues depending on how well they know the listener. P2 gave an example of using AAC with her doctor, while P1 typed a response but then indicated “no, I’m good” with a gesture, suggesting that even the decision not to speak can be intentional backchanneling in certain contexts.

Finally, participants shared what surprised them or what they learned. P3 stated, “I learned that there is some other new [AAC] technology out there that I want to try,” while P4 realized, “that my chair could help me communicate.” P2 offered a broader reflection: “People communicate in different ways, and that’s what makes us unique,” and P1—finishing her typed message—wrote, “I didn’t know what backchanneling was until today.” 

\section{Expert interviews: Backchanneling in Speech-Language Pathology}
To situate the workshop findings within common practice, we conducted follow‑up semi-structured interviews with four Speech‑Language Pathologists~(SLPs). 

\subsection{Participants}

Table~\ref{tab:slp-participants} lists our participants recruited through our network. They are all practicing SLPs.

\begin{table*} [t]
    \centering
    \resizebox{\textwidth}{!}{
    \begin{tabular}{lll}
        \textbf{ID} & \textbf{Clients} & \textbf{Experience} \\
        SLP1 & Emergent communicators with complex brain injuries & 6 years in specialized school \\
        SLP2 & Pediatric users (low and high tech) & 2.5 years in school and rehab clinic\\
        SLP3 & Mostly young children with mid to high tech SGDs and adult eye‑gaze users & 1 year in hospital and 2 externships\\
        SLP4 & Autistic and CP children and adults with motor impairments & 1 year in outpatient neuro-rehab \\
    \end{tabular}
    }
    \caption{Participants for our expert interviews.}
    \label{tab:slp-participants}
    \Description{
A table with a list of the participants in our expert-interview study. 
SLP1 has six years’ experience in a specialized school for children and adolescents with complex brain injuries, working mostly with emergent communicators. 
SLP2 has 2.5 years of experience, split her time between a school setting and, more recently, an outpatient rehab clinic, giving her exposure to both low‑tech pediatric users and high‑tech pediatric/adult clients.
SLP3 has 1 year of experience practicing with her license, plus 2 externships. She is a hospital‑based clinical fellow at the outpatient rehab setting, and also an elementary school sees mainly young children who use mid‑ to high‑tech speech‑generating devices, with occasional adult eye‑gaze cases.
SLP4 has 1 year of experience practicing in outpatient neuro‑rehab and an adult day‑habilitation program; she introduces AAC to autistic and CP children and customizes high‑tech systems for adults with motor impairments or post‑stroke aphasia.
    }
\end{table*}


\subsection{Data Collection and Analysis} 
All interviews were transcribed using Whisper's pre-trained large-v2 model~\cite{radford2023robust} on a local GPU (NVIDIA GeForce RTX 4080). After a confidence check of the transcripts by the interviewer, we used qualitative coding with 3 researchers (the first author and two external researchers) who focused their observations on the lived experience of the SLPs and the people they support and how they approach backchannelling for different AAC populations. Through iterative discussions, these observations were clustered into broader themes that captured key insights across participants. 

\subsection{Procedure}
Interviews took on average one hour and were conducted by the first author of this paper.  As mentioned in the positioning statement, the lead author of this work has a speech disability. Thus, interviews were conducted using his AAC device, which added a layer of observation to the conversation and interview process. Questions contained examples of observations from the workshop, and prompted about their general experience with backchanneling. The structure of the interview is available in the supplementary materials of this paper.

\subsection{Results}

While none of the four speech‑language pathologists used the specific term backchanneling in their clinical training or documentation, they were familiar with the underlying behaviors. These behaviors—such as nods, eyebrow raises, brief vocalizations, and shifts in eye contact—were typically described using broader clinical constructs like active-listening, non-verbal communication, pragmatic language skills, or prosody.

\subsubsection{Experience with AAC and backchanneling}
SLP1 and SLP2 described spending most of their time with “emergent” pediatric users whose first priorities are access, regulation, and words for basic survival (“food, water, bathroom, help”). In those contexts, conversational skills such as backchanneling are aspirational but secondary; the clinician’s immediate goal is to establish a reliable attention‑getter and survival messages. By contrast, SLP3 and SLP4 routinely support adults with different cognition (e.g., clients with progressive neurodegenerative disease) where the communicative ambition is full participation in multi‑turn dialogue. Here, backchanneling is framed as very important by SLP4, for sustaining flow and social belonging, yet is simultaneously constrained by motor abilities, device dwell‑time, and the need for partner training.

Every participant highlighted timing as the decisive hurdle. Partners must build in “expectant, inviting pauses”~(SLP1) because “AAC users just need a little bit more of… pause time”~(SLP2) before they can nod, vocalize, or hit a quick‑fire “yeah.” When that space is granted, users draw on a rich palette of idiosyncratic cues: “anywhere from an eyebrow raise to a tiny slight smile” in clients with limited mobility~(SLP4), a raised hand that says “give me one moment”~(SLP2), or the “intense or like intentional eye contact” that signals “I’m still composing”~(SLP2). These behaviors often substitute for the quick “mm‑hm” available to speaking partners, underscoring that effective backchanneling in AAC is co‑constructed through both device affordances and partner accommodation.

\subsubsection{Current Approaches and Training}
None of the clinicians has a formal “backchanneling curriculum,” but they weave the skill into therapy in three informal ways: noticing, modeling, and programming. SLP1 described an observation‑first stance: “I notice what they’re already doing… ‘Oh, you turned your head toward me.’ That tells me that you might be listening.” Rather than inventing new behaviors, she “shapes” the client’s existing cues so the learner realizes how others will interpret them. SLP3 echoed that most instruction is “more like implicit modeling... they imitate my hmm or my facial expressions.” Only when pragmatic goals are explicit (e.g., autistic children working on social skills) does SLP4 run direct drills on “facial expressions and nonverbal communication.”

Time constraints and clinical priorities often push formal teaching down the hierarchy. As SLP2 put it, “I’m still on that foundational step, just getting somebody to express needs for safety before building these other levels... I definitely wouldn’t say I directly work on that [backchanneling].” Yet even without explicit lessons, the therapists, who recognized these behaviors under different terminology, have their behaviors seep into sessions: “I’m sure anybody I’m working with is picking up on the cues I use all the time”.

When backchanneling phrases are needed, quick‑fire programming is the go‑to strategy. SLP4 routinely adds buttons like “hahaha,” “fist‑bump me,” or records a family member’s laugh because the default synthetic versions are “pretty bad… not very natural.” Still, most patients prefer faster, embodied signals. As SLP3 observed, “Even if ‘yes’ and ‘no’ are right there on the device, they’ll give me a head‑nod or thumbs‑up—it’s quicker.” In short, current practice favors recognizing and amplifying the AAC user’s existing multi-modal cues, with targeted programming only when a specific conversational shortcut is clearly valuable.

\subsubsection{Barriers and Challenges}
Common barriers and challenges that were noted by SLPs included timing constraints, misinterpretations of backchanneling behaviors, and language differences impacting communication. SLP2 emphasized the impact the environment has on AAC interactions, in response to seeing a clip of our co-design workshop: "I think if you're thinking about the communication environment that that group was held in, it's a little bit different than the communication environment that I have because I'm one-on-one, and I think that in a group setting, AAC is definitely more challenging." Depending on the noise level of the environment, it can distract communication partners, making it more difficult for AAC users to keep up with the conversation. SLP1 had also mentioned the impact of the environment when saying, "We can't always rely that everyone's going to be in an environment where they have someone just like baseline paying attention". Body positioning was another variable that impacted AAC interaction, which was noted by SLP3:~"The tips for having a good conversation with someone I always say, make sure you're sitting right in front of them". Having communication partners sit in front of one another in a quiet environment allows AAC users to quickly interpret and respond to backchanneling behaviors, reducing the need for clarification or repetition. 

Although timing constraints were expressed by some participants, there was overlap between timing and misinterpretation. Sometimes slow or delayed motor movements can impact communicative intent as mentioned by SLP3:~"sometimes I can only imagine like how much harder it is to then like try to go all the way back when we're already kind of moved on to a different topic." SLP2 had also noted missed opportunities where communication partners don't notice when the AAC user is using nonverbal cues, such as when they are "agreeing or disagreeing": "a lot of times, the immediate yes or no, I think is the most one that I've missed.". 

Other backchanneling behaviors that were misinterpreted included vocalizations. For instance, SLP3 had mentioned one of her clients "verbalizing an 'eee' sound quite frequently"; however, it was unclear what their communicative intent was. This further emphasizes the uniqueness of preferred communication styles used by AAC users; they will produce vocalizations which are "most natural for them to produce."~(SLP4). Similar to motor movements as expressed by SLP2:~"for people with limited motor abilities, I would say it [their motor movements] looks different depending on the person." SLP2 saw difficulties in using motor movement to "interject" into the conversation, as "it's hard for people to understand their interjection and what they mean by it". SLP2 had also emphasized that motor impairments can create difficulty for clarifying their communicative intent because "sometimes it's hard to locate the right button [using touch screen or gaze]". 

Language differences were observed to be a challenge, as SLP1 indicated that one of her clients switches between two languages during a conversation. SLP1 indicated that communication partners must be aware of those who may speak another language and give them space to join in on the conversation.

\subsubsection{Device Features and Techniques}  

When discussing device features and techniques, all participants noted device modifications for increased client engagement. For SLP1, “programming is everything.” \textit{Programming} in the context of AAC technology is mapping pre-defined utterances to keys on the speech generating device, a common feature to do this is called \textit{QuickFires}. Her students thrive when acknowledgments sit on a single \textit{QuickFires} page: “One of my kids is a super big \textit{QuickFires} page [user] and they love knowing, like, ‘oh my gosh, I have all these really quick responses – yes, no, just kidding, stop that, get out of here, let me see that – all on one page.’”~(SLP1). Because “timing is most relevant for these kinds of phrases,” she co‑creates the content: “We’re spending time figuring out what words, what phrases, how you would like to be backchanneling, and we’re giving you ways to do that.”~(SLP1). SLP1 also provides her clients with increased opportunity for backchanneling by encouraging use of their built-in \textit{QuickFires} page as it contains universal phrases such as "yes", "no" and "stop that".

SLP4 also relies on built‑ins but will reposition them for speed:  
“The device usually comes with settings that are pre‑typed… If the user wants [a phrase] more accessible, I make sure it’s on the home screen and in a place motor‑wise where they can get to it the quickest.”~(SLP4)  
“I think the ‘I’m typing’ one is good, just a \textit{QuickFire} – even yes/no is good.” In addition to pre-programmed phrases, SLP2 noted one client who insists on a single button that says “I’m listening” because he “loves to just constantly be engaged and let others know that he is there.” Over time, she learned to read his gestures: “Now we have such a better relationship… I know when he uses a certain gesture when he’s actually telling me he has something to say versus when it’s just his way to engage – almost like his 'mm‑hm'.”~(SLP2)  
  
SLP3 takes it further by rearranging vocabulary so it can be reached mid‑conversation: “I adjust navigation, moving frequently used words or phrases so the AAC user can navigate to them more efficiently. I find that they’re used more.”~(SLP3). But for backchanneling, still default to embodied cues: “\textit{TouchChat} has a ‘that’s cool’ phrase, but for my older adults we usually go back to a vocal or a gesture or eye contact.”~(SLP3). All participants stressed that bodies beat buttons for real‑time acknowledgments: “Nonverbal communication modalities are extremely, extremely, extremely important… a facial expression or vocalization is quicker than generating a message response.”~(SLP2). “I always am preaching multi-modal communication because, at the end of the day, the device is not the person talking… look at the user and not the device.”~(SLP4). Other modes of communication, such as gestures or vocalizations, allow AAC users to express backchanneling behaviors without experiencing time constraints.  

Together, these accounts show that while custom shortcuts and layout hacks help, successful backchanneling hinges on a weave of \textit{QuickFire} phrases, clever positioning, and the AAC user’s own multi-modal repertoire.

\subsubsection{Balancing Diverse Needs Without Imposing One ‘Norm’}
Participants see backchanneling as a culturally negotiated practice that should remain fluid rather than standardized. SLP3 doubted a universal scheme could ever fit: “That single universal approach… I wonder if that’s even possible… there’s just so much variation.” She added that she would “be interested to hear from AAC users from around the country, around the globe…I think of sign language as an example of like different dialects... how it is different in different countries and different individuals.” SLP4 embraced the idea of an AAC “micro-culture”, affirming: “I love that word, micro-culture... any kind of single universal approach to AAC is pretty bad because it’s just so different working with different people.”

To let mainstream norms and AAC micro‑cultures coexist, participants called for design strategies that foreground user agency and partner education rather than one fixed template. SLP1’s method is to surface and legitimize each person’s natural cues: “I notice what they’re already doing… accept that and teach you and everyone around you that this is how you do it.” SLP2 framed the broader social task as cultivating empathy: “We need… more conversations on these topics and more exposure in order to shift people to having a little bit more empathy and acceptance” of different "backchanneling dialects", just as society is learning to respect dialectal differences in spoken language.

Technology should therefore expand, not narrow, expressive possibilities. As SLP4 noted, “there's so much room for personalization. So I see that as one of my main jobs when I work with people, they see how I make this[AAC], their voice?” Coupled with timing controls that let users match the pace of diverse partners—“Timing would be a really, really great… customization feature”~(SLP2)—such flexibility allows AAC users to uphold their own conversational agreements while still engaging smoothly with the wider communicative culture.

\subsubsection{Opportunities for Backchanneling in Training and Technology}
Participants agreed that backchanneling remains a blind spot in both AAC software and therapy. SLP2 captured the sentiment bluntly: “I think the whole topic of backchanneling with AAC is under‑explored in general… It’s often overshadowed by ‘let’s get those sentences out.’”

On the technology side, SLP1 pointed to emerging features that could be harnessed but rarely are: “You can play recorded videos or play recorded speech, you could get a sibling’s voice, even a sibling’s facial expression,” Such multi-modal triggers, she argued, offer a fast, personalized way to acknowledge a partner.

SLP4 sees untapped potential inside mainstream AAC apps: “Being able to access these [backchanneling cues] in their own sidebar, making the voices sound more natural and more variable based on emotions… maybe even adding some kind of gestures on there if people aren’t able to produce their own.” She called the entire area “so much room for improvement.” 

Training and professional awareness are part of the problem. SLP2 confessed, “I don’t think I ever truly, in my practice or even in my education, remember a time being explicitly taught to approach this… until just now I’m like oh wait, these are important, like super‑important things.” she added "also for non-AAC user, kind of these things that are almost so ingrained in us that we are not always consciously thinking about these subtle things that we're doing, where for an AAC user, that might not necessarily be the case." Her realization was echoed by SLP4—“I would say all the strategies are under‑explored. I truly do not believe that this is talked about enough, and I do agree that it's pretty important, especially during communications breakdowns". 

Together, they highlight a systemic gap: backchanneling rarely appears in coursework, clinical guidelines, or software, signaling an urgent need to elevate it into standard curricula and therapy protocols. These conversations prompted SLPs to reflect more deeply on the importance and uniqueness of backchanneling for AAC users and acknowledged that it remains underrepresented in professional education.

\section{Discussion}
In this section, we combine insights from our studies. We consider these starting points for further exploration rather than broadly generalizable "facts," given the small and specific study population.

\subsection{Backchanneling as Micro‑Cultural Practice}
Our studies showed how AAC users form their own "listener-feedback dialect." Participants routinely mixed device output with eyebrow raises, vocalizations, wheelchair taps, and partner‑assisted paraphrases. These practices form a micro‑culture—a locally negotiated style of engagement that intersects with and diverges from mainstream norms ~\cite{fogel1993developing}. Crucially, this culture shifts with partner composition: AAC–AAC pairs struggled with missed visual cues when both looked down to type, whereas AAC–non‑AAC pairs used repetition (“You said poetry?”) to maintain shared understanding.

Backchanneling is culturally negotiated—what counts as an encouraging listener cue in the United States (e.g., a rapid “yeah‑yeah”) is not identical to the overlapping “eh” or emphatic hand gestures common in Italy, or the "aizuchi" (e.g., “hai,” “sou desu ne”) preferred in Japan. A striking example of a micro-culture for backchanneling is found in the Deaf-blind community’s use of \textit{Protactile Language}, a fully touch-based language that redefines listener feedback through taps, squeezes, and shared contact space~\cite{granda2018protactile}. 

If we treat AAC communication as its own micro‑culture with ‘rules of engagement,’ then cross‑cultural conversation (with non-AAC users) becomes a meeting of dialects. Building on \citet{hubscher2023role}, who conducted a multi-modal analysis of discourse markers in AAC interviews, our findings suggest treating AAC communication as a micro‑culture that foregrounds the need for design approaches that expose rather than overwrite situated practices. This approach empowers the AAC community by claiming the uniqueness of each individual’s self-expression, a key issue found by others studying expressive AAC communication~\cite{kane_at_2017,valencia_less_2023,weinberg2024why}. When we consider this a cross-cultural problem, we see that backchanneling requires mutual adaptation and respect for each user’s timing, modality, and micro-cultural norms, rather than imposing mainstream standards.

\subsection{Rethinking Embodiment and Mediation in AAC Technology}
Most AAC technologies treat the body as an input mechanism for transmitting digital messages—a conduit to the “real” act of communication, which happens via a screen and speech synthesis. This reflects a \textit{transmission model of communication}~\cite{ellis1990if}, where meaning flows linearly from sender to receiver through a technological channel. Under this framing, backchanneling becomes a challenge of latency: how quickly can the system convert intention into output?

In our findings, the meaning of backchanneling was co-constructed through shared context and reciprocal cues. Both partners are simultaneously senders and receivers, continually communicating their presence even as the other is sending an explicit message~\cite{barnlund1970transactional, waller2006communication}. The resulting co-creation of meaning by both partners impacts the flow of the conversation. Whether it is finding meaning through movement or vocalization, it is important how each partner interprets the signals from each other. Technology is not the default medium but serves as a fallback or enhancement when bodily signaling is unavailable or insufficient. \citet{valencia_conversational_2020} emphasized how responsiveness and timing from communication partners can scaffold or hinder an AAC user’s ability to maintain control in a conversation. Our findings further support this by illustrating how backchanneling serves as a mechanism through which agency is negotiated in real time.

This perspective has implications for designing AAC technology. We ask: how can we foreground the human body as the primary communicative site, and invoke technology only when necessary? This framing treats the body as the baseline and technological mediation as a situational augmentation.

Designing with this perspective has several implications:
AAC systems could detect and respond to nonverbal feedback cues like facial expressions, posture changes, or even physiological signals. These embodied signals can convey attention, affect, or intent more naturally than device-generated speech. Rather than focusing solely on transmitting content, interfaces such as COMPA~\cite{valencia2024compa} can support turn-taking and mutual attunement—e.g., showing “typing,” “thinking,” or “still here” states through ambient cues without compromising privacy~\cite{gonzalez2025trusting}.

This re-framing pushes us toward human-centered AAC design~\cite{curtis2024beyond}, where backchanneling is not about embodying the virtual, but about augmenting the body’s innate communicative capacities when needed. Rather than optimizing AAC for faster transmission, we might instead focus on better participation, anchored in the body, and mediated only when that body asks for support.

\subsection{Integrate Backchanneling in SLP curricula}
We identify a crucial disconnect between the social need and desire for backchanneling and the way SLPs are educated and how they teach their clients. Both workshop and interview participants highlight the importance of backchanneling, albeit using different terminology such as active listening. However, none were explicitly educated or provided training to their clients on this topic. While technology can help mediate backchanneling, in line with observations by \citet{buzolich1988turn} in the 80s, we identify a societal opportunity to bring backchanneling techniques into education, both to educate SLPs and for SLPs to teach their clients. We hope that the observations presented in this paper serve as a starting point for this inclusion.

\subsection{Designing for Timely Multi-modal Backchanneling}

Designing AAC tools for expressive backchanneling demands a shift in emphasis from composing speech to communicating presence. \citet{kane_at_2017} underscores the importance of self-expression in AAC use, revealing how individuals with ALS adapt language and timing to maintain a sense of personality. Building on this, our findings show that timely listener feedback often occurs not through words, but through coordinated bodily, auditory, and environmental signals—many of which operate outside traditional AAC interfaces. Participants in our workshop repeatedly relied on subtle gestures (e.g., eyebrow raises, head nods), device workarounds (e.g., \textit{QuickFires}), and even chair movements (e.g., “beeps,” repositioning) to signal engagement. These strategies served as parallel channels, quick, expressive, and less disruptive to the message composition. 

AAC tools should support communicating presence, allowing users to express attention and understanding while composing a message. For instance, a raised eyebrow might signal “I'm following,” even as the user types a longer response. Such nonverbal cues, though idiosyncratic, are often more immediate than any device-mediated utterance.

In practical terms, this points to three design opportunities:
\begin{itemize}
    \item \textbf{Fast, low-effort cues:} participants defaulted to physical expressions when verbal output took too long. \textit{QuickFire phrases} like “wait,” “yeah,” or “I’m listening” were helpful, but constrained by navigation time and device layout. Future AAC interfaces could offer multi-modal feedback triggers (e.g., eyebrow-detection, physiological sensors, or vibration cues) that complement typing without interrupting it.
    \item \textbf{Real-time customizable, expressive shortcuts:} Participants wanted to control both the content and tone of their backchanneling—e.g., a supportive “wooow” vs. a skeptical “really?” Rather than one-size-fits-all templates, AAC systems should enable users to create rich, emotion-tagged presets (e.g., “surprised mm-hmm” or "agreeing hmmhm") that align with their personal style and context or adjust the tone on the fly.
    \item \textbf{Partner-aware timing support:} Backchanneling was shaped by who the user was speaking with. In AAC–AAC interactions, missed visual cues and overlapping typing led to delays or topic mismatches. In line with ~\cite{bloch_understandability_2004, tegler_aided-speaking_2024} work, which explored strategies like rephrasing, pre-playing utterances, or relying on partners to scaffold the interaction to manage turn-taking asymmetries, we found that in AAC–non–AAC settings, repetition and physical prompts helped bridge the pace mismatch. Interfaces could assist by visualizing floor dynamics (e.g., “your partner is typing,” or “message received”), or offering predictive timing aids (e.g., a “hold on” gesture while composing).
\end{itemize}

Ultimately, the challenge is not just to accelerate backchanneling but to re-center it as a socially meaningful, culturally shaped practice. A well-designed AAC system should empower users to say “mm-hmm” in their own way—whether through a blink, a buzz, or a burst of sound—on their terms, and in their timing. As our participants made clear, backchanneling isn’t about finishing someone’s sentence. It’s about holding space for theirs.

\subsection{Future Work}
These findings are based on qualitative data from two relatively small-scale controlled studies. While the results serve as starting points or inspiration for further research, they cannot be considered \textit{generalizable}, specifically, given our own observation that backchanneling is cultural and locally co-created instead of universal and generic. Future work could revisit these studies with a broader population or in a more natural setting. 

Our contextualization with SLPs was helpful to expand our findings beyond the specific workshop setting and learn from long-term observations. It would still be interesting to see follow-up studies to capture a first-hand perspective of longer-term backchanneling. 
\section{Conclusion}
In this paper, we presented an exploration of backchanneling in the AAC \textit{micro‑culture}, examining how AAC users navigate “timing, effort, and modality” to produce listener feedback that is as expressive and socially binding as direct backchanneling in typical speech. Through a co‑design workshop with four AAC users and follow‑up interviews with four speech‑language pathologists, we surfaced the multi-modal strategies that AAC users and their partners co‑construct to keep conversation flowing. Building on these insights, we proposed design recommendations that treat backchanneling as a first‑class goal when developing new AAC technology: fast and customizable timing controls paired with partner‑training scaffolds that respect diverse “cultural agreements” rather than imposing mainstream norms. Besides technological implications, our findings also point to a gap in SLP curricula to help AAC users take advantage of this rich form of multi-modal communication. 

We see this as a small step to better understand expressive communication for AAC users. Through such expressive, informal communication that people connect on a deeper level. Understanding how AAC users navigate these communication barriers is crucial to develop technology that enables them to fully flourish in society.

\begin{acks}
We extend our sincere gratitude to YAI (Young Adult Institute) for providing the space and facilities that enabled this research. We are especially thankful to Judith Baley-Hung, BIS Coordinator and Supervisor of the Center for Engagement and Innovation at YAI, for her mentorship and support throughout the project. We also thank Julia Shuman for connecting us with a generous and insightful network of speech-language pathologists. This work was supported by the Siegel Public Interest Technology Impact Fellowship (PiTech), whose funding made this research possible.
\end{acks}

\bibliographystyle{ACM-Reference-Format}
\bibliography{ASSETS2025}


\end{document}